\documentclass[12 pt] {article}
\begin{document}
\begin{center}


{\bf Chaotic oscillations in nonlinear system of interacting
oscillators with the interaction of the fourth order.}

\vspace{0.8cm} {L. Chotorlishvili
  }

\vspace{0.5cm}\baselineskip=14pt

\vspace{0.5cm} \baselineskip=14pt {\it Tbilisi State University,
Georgia, 0128. Tbilisi, \\ Chavchavadze av. 3, Email:
lchotor@yahoo.com
 }
\end{center}
\vspace{0.5cm}

\begin{abstract}
 Dynamics of two anharmonic oscillators with interaction of the
fourth order has been investigated. The conditions at realization
of which system is integrable are established. The exact
analytical solution of the nonlinear equations in the case of
adiabatic isolation of a system of oscillators has been obtained.
\end {abstract}

\par
In optics of weak light streams the polarization of an electromagnetic wave
in substance does not depend on intensity of light and univocally
corresponds to the polarization of emanation, impinging on the boundary of
medium - vacuum. The situation cardinally varies in the case of nonlinear
optics, when the factors of a refraction and absorption of substance become
functions of intensity of a radiation [1]. At a particular combination of
parameters of a radiation and nonlinear medium, the stationary polarization
of light '' on an inlet '' becomes unstable '' on an exit '', or there can
be a random modification of polarization in time [2-7]. The specified
phenomenon is called as a spontaneous violation of polarization symmetry.

In a strong light field, when the nonlinear component of the electrical
induction becomes noticeable with respect to the linear component,
traditional exposition based on a material equation of a medium, which
represents a series with respect of terms of electric field is impossible.
In works [8-9] the prospects of observation of amplitude instability for
mediums in which nonlinear response is featured within the framework of a
model of the Duffing oscillator were considered. In work [10] the nonlinear
susceptibility of the ensemble of chaotically oriented mirror - asymmetrical
molecules has been estimated. Assumption that each of molecules can be
represented by the nonlinear system of interacting oscillators in the field
of electromagnetic wave has been done:
\begin{eqnarray} \ddot x+\omega _{0}^{2}x+\alpha y+\Gamma \dot x+ax^{3}+b(y^{3}+3x^{2}y)+cxy^{2}=
\frac{eE_{z}}{m}\exp(iK_{z}\frac{D}{2}) \nonumber
\\ \ddot x+\omega _{0}^{2}y+\alpha x+\Gamma \dot y+ay^{3}+b(x^{3}+3y^{2}x)+cyx^{2}=\frac{eE_{y}}{m}\exp (iK_{z}\frac{D}{2}%
) \end{eqnarray}

Where $\omega _{0}\,\,$is a natural frequency of small
oscillations of oscillators creating a molecule, $ e $ and $ m $
electrical charge and mass accordingly, $E_{z}$, $E_{y}\,\,$
projections of amplitude of an electric field, $ D $ is a distance
between oscillators, $ a $ , $ b $ , $ c
$ are the anharmonic
constants naturally of oscillators and connection between them,
$\Gamma \,$\ is the parameter describing attenuation. With the
help of summation of the dipole moment of molecules

 $$ d_{x,y}=e\left( \begin{array}{c} x \\ y \end{array} \right) \exp (\pm iK_{z}\frac{D}{2}) $$

and averaging with respect of small volume, in the work [10], the expression
for macroscopic polarization was obtained, and was shown that in mediums
with nonlinear gyrotropy the polarization instability takes place.

Despite of it, due to the great importance for the nonlinear optics, the
immediate research of the system of nonlinear oscillators (1) is of interest.

The purpose of given work is the qualitative, theoretical research of the
equations (1) and establishment of relations between parameters of a system
at which equation (1) will be integrable.

Let us consider at first the case of weak nonlinearity. We also suppose that
the amplitude of the external field is small:

 $$ \omega _{0}^{2}>>\frac{eE_{z}}{m} $$

Considering terms with cubic nonlinearity as a perturbation, let
us rewrite system (1) in following form:
\begin{eqnarray} \ddot x +\omega _{0} ^{2} x +\alpha y=-\mu f(x,\dot x, y,\dot y)
,\nonumber \\ \ddot y +\omega _{0} ^{2} y+\alpha x=-\mu g(x,\dot
x, y,\dot y), \end{eqnarray}

where
\begin{eqnarray}
f(x,\dot x,y,\dot
y)=x^{3}+y^{3}+3x^{2}y+xy^{2}+\frac{\Gamma}{\mu}\dot x \nonumber
\\ g(x,\dot x, y, \dot y)=x^{3}+y^{3}+3xy^{2}+x^{2}y+\frac{\Gamma}{\mu}\dot y
\end{eqnarray}

here are used the following designations:

$$ x\rightarrow a_{0}x, \, y\rightarrow a_{0}y, \, \omega_{0}\rightarrow\omega_{0}
\sqrt{\frac{ma_{0}}{eE_{z,\,y}}}\,,\,\Gamma\rightarrow\Gamma\sqrt{\frac{ma_{0}}{eE_{z,\,y}}}\,,\,
\alpha\rightarrow\alpha\frac{ma_{0}}{eE_{z,\,y}} \,, t\rightarrow
t \sqrt{\frac{eE_{z,\,y}}{ma_{0}}}$$
$$ \mu\approx a\frac{ma_{0}^{3}}{eE_{z,\,y}}\approx b\frac{ma_{0}^{3}}{eE_{z,\,y}}\approx
c\frac{ma_{0}^{3}}{eE_{z,\,y}}, $$ and transformation to the
dimensionless variables and parameters (where - $ a_{_{0}}\,$
means a Bohr radius) has been done. For $\mu =0$ the set of
equations (2) has the solution: \begin{eqnarray} x=\alpha
_{_{1}}\sin (k_{_{1}}t+\beta _{_{1}})+\alpha
_{_{2}}\sin (k_{_{2}}t+\beta _{_{2}}) \nonumber \\
y=-\alpha _{_{1}}\sin (k_{_{1}}t+\beta _{_{1}})+\alpha _{_{2}}\sin
(k_{_{2}}t+\beta _{_{2}}) \end{eqnarray}  Here $\alpha
_{_{1}},\,\alpha _{_{2}}, \,\beta _{_{1}}, \,\beta _{_{2}}\,$ are
the constants of integration, \ $k_{_{1}}=\sqrt{\omega
_{0}^{2}-\alpha _{_{1}}},\,\ k_{_{2}}=\sqrt{\omega _{0}^{2}-\alpha
_{_{2}}}$ \par Solution of the set of equations (2) for $\mu\neq
0$  we shall search in the form (4), considering that $\alpha
_{_{1}}\,,\alpha _{_{2}}\, , \beta _{_{1}}\, ,\beta _{_{2}}$ are
slowly varying functions of time. The additional conditions which
we shall impose on the functions $\alpha _{_{1}}\, ,\alpha
_{_{2}}\, , \beta _{_{1}}\, , \beta _{_{2}}$ for their
determinancy consist in the following: first order temporary
derivatives of $ x $ and $ y $  are having the same form, as in
the case of constant $\alpha _{_{1}}\, ,\alpha _{_{2}}\, , \beta
_{_{1}}\, , \beta _{_{2}}$. As a result we get:\begin{eqnarray}
\dot \alpha_{_{1}} \,sin(k_{_{1}}t+\beta_{_{1}})+\dot
\alpha_{_{2}}
\,\sin(k_{_{2}}t+\beta_{_{2}})+\alpha_{_{1}}\dot\beta_{_{1}} \,
\cos(k_{_{1}}t+\beta_{_{1}})+\alpha_{_{2}}\dot\beta_{_{2}}\,
\cos(k_{_{2}}t+\beta_{_{2}})=0 \nonumber \\ -\dot \alpha_{_{1}}
\,\sin(k_{_{1}}t+\beta_{_{1}})+\dot \alpha_{_{2}}
\,\sin(k_{_{2}}t+\beta_{_{2}})-\alpha_{_{1}}\dot\beta_{_{1}} \,
\cos(k_{_{1}}t+\beta_{_{1}})+\alpha_{_{2}}\dot\beta_{_{2}}\,
\cos(k_{_{2}}t+\beta_{_{2}})=0 \end{eqnarray} \par Calculating
temporary derivatives of $ \dot x,$ \, and $\dot y \,\,$and
substituting obtained expressions in the equations (2) one can
get:
\begin{eqnarray} \dot \alpha_{_{1}}k_{_{1}}\cos(k_{_{1}}t+\beta_{_{1}})+
 \dot \alpha_{_{2}}k_{_{2}}\cos(k_{_{2}}t+\beta_{_{2}})-k_{_{1}}\alpha_{_{1}}\dot\beta_{_{1}}
 \,\sin(k_{_{1}}t+\beta_{_{1}})+k_{_{2}}\alpha_{_{2}}\dot\beta_{_{2}}
 \,\sin(k_{_{2}}t+\beta_{_{2}})=\mu f{\ast} \nonumber \\ -\dot \alpha_{_{1}}k_{_{1}}cos(k_{_{1}}t+\beta_{_{1}})+
 \dot \alpha_{_{2}}k_{_{2}}\cos(k_{_{2}}t+\beta_{_{2}})-k_{_{1}}\alpha_{_{1}}\dot\beta_{_{1}}
 \,\sin(k_{_{1}}t+\beta_{_{1}})-k_{_{2}}\alpha_{_{2}}\dot\beta_{_{2}}
 \,\sin(k_{_{2}}t+\beta_{_{2}})=\mu g{\ast} \end{eqnarray} Where
\begin{eqnarray}
f^{\ast }=f[\alpha _{1}\sin (k_{1}t+\beta _{1})+\alpha _{2}\sin
(k_{2}t+\beta _{2})\,;\alpha _{1}k_{1}\cos (k_{1}t+\beta
_{1})+\alpha _{2}k_{2}\cos (k_{2}t+\beta _{2})\,; \nonumber \\
-\alpha _{1}\sin (k_{1}t+\beta _{1})+\alpha _{2}\sin (k_{2}t+\beta
_{2})\,;\,-\alpha _{1}k_{1}\cos (k_{1}t+\beta _{1})+\alpha
_{2}k_{2}\cos (k_{2}t+\beta _{2})\,\,],\nonumber \\
g^{\ast }=g[\alpha _{1}\sin (k_{1}t+\beta _{1})+\alpha _{2}\sin
(k_{2}t+\beta _{2})\,;\alpha _{1}k_{1}\cos (k_{1}t+\beta
_{1})+\alpha _{2}k_{2}\cos (k_{2}t+\beta _{2})\,; \\
-\alpha _{1}\sin (k_{1}t+\beta _{1})+\alpha _{2}\sin (k_{2}t+\beta
_{2})\,;\,-\alpha _{1}k_{1}\cos (k_{1}t+\beta _{1})+\alpha
_{2}k_{2}\cos (k_{2}t+\beta _{2})\,\,] \nonumber
\end{eqnarray}

The equations (5), (6), (7) are a set of equations for the
determination of $ \alpha _{_{1}} \, ,\alpha _{_{2}} \, , \beta
_{_{1}}\, , \beta _{_{2}}$.

After solving of these equations with respect to $
\dot\alpha_{_{1}}$one can obtain:
 $$\frac{d\alpha _{1}}{dt}=\frac{\Delta _{1}}{\Delta} $$
$$
\Delta=\left|\begin{array}{clcr}\sin\xi &\sin\eta &\cos\xi&\cos\eta \\
          -\sin\xi & \sin\eta & -\cos\xi &\cos\eta\\
          k_{1}\cos\xi & k_{2}\cos\eta & -k_{1}\sin\xi
          & -k_{2}sin\eta\\
          -k_{1}\cos\xi & k_{2}\cos\eta & k_{1}\sin\xi &
          -k_{2}\sin\eta\end{array}\right| ,$$

$$
\Delta_{1}=\left|\begin{array}{clcr} 0 &\sin\eta &\cos\xi&\cos\eta \\
          0 & \sin\eta & -\cos\xi &\cos\eta\\
          \mu f^{\ast} & k_{2}\cos\eta & -k_{1}\sin\xi
          & -k_{2}sin\eta\\
          \mu g{\ast} & k_{2}\cos\eta & k_{1}\sin\xi &
          -k_{2}\sin\eta\end{array}\right| ,$$

\bigskip here $\xi =k_{1}t+\beta _{1}$and $\eta =k_{2}t+\beta _{2}$.

Making similar operations with respect to $\alpha _{1} \, ,\alpha
_{2} \, ,  \beta _{1}\, , \beta _{2}$ after unwieldy calculations
we get:\begin{eqnarray} \frac{d\alpha _{1}}{dt}=\frac{\mu }{
k_{1}(k_{2}^{2}-k_{1}^{2})}[\alpha f^{\ast }-\alpha g^{\ast
}]\,\cos \xi ,\nonumber \\ \frac{d\alpha _{2}}{dt}=\frac{\mu }{
k_{2}(k_{2}^{2}-k_{1}^{2})}[\alpha f^{\ast }+\alpha g^{\ast
}]\,\cos \eta ,\nonumber \\ \alpha _{1}\frac{d\beta_{1}}{dt}
=\frac{ \mu }{k_{1}(k_{2}^{2}-k_{1}^{2})}[\alpha f^{\ast }-\alpha
g^{\ast }]\,\sin \xi ,\\
\alpha_{2}\frac{d\beta_{2}}{dt}=-\frac{\mu
}{k_{1}(k_{2}^{2}-k_{1}^{2})}[\alpha f^{\ast }+\alpha g^{\ast
}]\,\sin \eta . \nonumber
\end{eqnarray}

The obtained equations (8) represent a set of equations (2),
transformed to another variables. Let's assume, variation of the
$\alpha _{1}\, ,\alpha _{2}\, , \beta _{1}\, , \beta _{2}$ are
slow than oscillations in the initial dynamical system. Averaging
the obtained equations with respect to phase $ \frac{2\pi
}{k_{1}}$ and $\frac{2\pi }{k_{2}}$ for $\alpha _{1}\, ,\alpha
_{2} \, , \beta _{1}\, , \beta _{2}$ we shall get:
\begin{eqnarray}
\dot\alpha _{1}=\frac{ \mu }{2\alpha \sqrt{\omega _{0}^{2}-\alpha
}}(F_{1}-G_{1}),\nonumber \\\dot\alpha _{2}=\frac{ \mu }{2\alpha
\sqrt{\omega _{0}^{2}+\alpha}}(F_{2}+G_{2}),\nonumber \\
 \dot\beta _{1}=-\frac{\mu }{2\alpha \sqrt{\omega _{0}^{2}-\alpha }}(F_{3}-G_{3}),\\
 \dot\beta _{2}=-\frac{\mu }{2\alpha \sqrt{\omega _{0}^{2}+\alpha
}}(F_{4}+G_{4}).\nonumber \end{eqnarray}

Here
\begin{eqnarray}
 F_{1}=\frac{1}{2\pi ^{2}} \int_{0}^{2\pi }\int_{0}^{2\pi
}f^{\ast }\cos \xi d\xi d\eta ;  G_{1}= \frac{1}{2\pi
^{2}}\int_{0}^{2\pi }\int_{0}^{2\pi }g^{\ast }\cos \xi d\xi
d\eta; \nonumber \\
 F_{2}=\frac{1}{2\pi ^{2}} \int_{0}^{2\pi }\int_{0}^{2\pi
}f^{\ast }\cos \eta d\xi d\eta ;  G_{2}= \frac{1}{2\pi
^{2}}\int_{0}^{2\pi }\int_{0}^{2\pi }g^{\ast }\cos \eta d\xi d\eta
; \nonumber \\
 F_{3}=\frac{1}{2\pi ^{2}}\int_{0}^{2\pi }\int_{0}^{2\pi }f^{\ast
}\sin \xi d\xi d\eta ;   G_{3}= \frac{1}{2\pi ^{2}}\int_{0}^{2\pi
}\int_{0}^{2\pi }g^{\ast }\sin \xi d\xi d\eta; \\
 F_{4}=\frac{1}{2\pi ^{2}} \int_{0}^{2\pi }\int_{0}^{2\pi
}f^{\ast }\sin \eta d\xi d\eta ;  G_{4}= \frac{1}{2\pi
^{2}}\int_{0}^{2\pi }\int_{0}^{2\pi }g^{\ast }\sin \eta d\xi d\eta
; \nonumber
\end{eqnarray}
The right sides of the equations (8) do not depend on $\beta _{1}$
and $\beta _{2}$. Therefore they serve for the investigation of
$\alpha _{1}$ and $ \alpha _{2}$. Substituting (7) in (10) and
integrating finally we get:
\begin {equation}
 \alpha _{1}(t)=\alpha _{1}(0)e^{-%
\frac{\Gamma }{\alpha (\omega _{0}^{2}-\alpha )^{1/2}}t}\,\,,\,\,\
\ \alpha _{1}(t)=\alpha _{1}(0)e^{-\frac{\Gamma }{\alpha (\omega
_{0}^{2}+\alpha )^{1/2}}t}\,.
\end {equation}
It is easy to see from (11) that in case of weak nonlinearity, the nonlinear
terms does not affect on the form of motion. Therefore the application of
the method of slowly varying amplitudes in case of strong nonlinearity made
in [9] is incorrect. As a result in the system originates damping
oscillations and the phase diagram represents a steady focal point. This
outcome can be considered as a singularity of the system (1). These
reasonings are proved also by numerical calculations (see Fig. 1).

Fig. 1 Phase diagram obtained by numerical calculations for the
values of the dimensionless parameters:\, $ \omega
_{0}^{2}=10\,,\,\ \alpha =1,\,\ \Gamma =0,5\,,\,\,f=0,1\,,\,\
a=0,6\,,\,\,\ b=0,5\,,\,\ c=0,5 $

As show numerical calculations at the action on the system of a strong
exterior field
\begin{equation}
 \omega _{0}^{2}=0.1\frac{eE_{x,\,y}}{ma_{0}}
 \end{equation}

in the phase diagram there is a limit cycle [11-13] (see Fig. 2).

Fig. 2 Phase diagram obtained by numerical calculations for the values of
the dimensionless parameters:

$\ \ \ \ \ \ \ \omega _{0}^{2}=10,\alpha =1,\,\Gamma =0.5,\,\
f=1.7\,,\,\,a=0.6\,\ \ b=0,5\,,\ \ c=0.5.$

In case of strong nonlinearity the application of the method of slowly
varying amplitudes becomes impossible. However below will be shown that
under particular conditions is possible to obtain analytical solutions of
the set of equations (1) even in the presence of interaction terms of the
fourth order.

It is easy to see that in case of absence external field and damping, (i.e.
in case of adiabatic apartness) the equations (1) can be generated by the
Hamiltonian
\begin{equation}
H=P_{x}^{2}/2+P_{y}^{2}/2+\omega _{0}^{2}/2\,(x^{2}+y^{2})+\alpha
xy+a/4(x^{4}+y^{4})+c/2\,x^{2}y^{2}+b(xy^{3}+x^{3}y)
\end{equation}

After the transformation to the variables $q_{1}=x+y\, ,
q_{2}=x-y$ and supposing that $c=3a\,$ one can obtain:
\begin{eqnarray}
H=H_{1}+H_{2}, \nonumber \\
 H_{1}=\frac{p_{1}^{2}}{2}+\frac{a+b}{8}q_{1}^{4}+\frac{1}{4}(\omega_{0}^{2}+\alpha)q_{1}^{2}, \\
H_{2}=\frac{p_{2}^{2}}{2}+\frac{a+b}{8}q_{2}^{4}+\frac{1}{4}(\omega_{0}^{2}-\alpha)q_{2}^{2}.\nonumber
\end{eqnarray}
The similar result can be obtained with the using of more general method
(the method offered by P.Lax [14]). This method allows solving the problem
even in case when determinations of new variables are not so trivial.

Let's initiate the analysis of the Hamiltonian (13). The corresponding
equations look like
\begin{equation}
\ddot q_{_{1,2}}=-\frac{a+b}{2}\,q_{_{1,2}}^{3}-\frac{\omega
_{0}^{2}\pm \alpha}{2}\,q_{_{1,2}}
\end{equation}

Representing $\ddot q_{_{1,2}}\,$ as $ \,\ddot
q_{_{1,\,2}}=\frac{1}{2\dot q_{_{1,\,2}}}\frac{d}{dt}(\dot
q_{_{1,\,2}}
)^{2} $and substituting in (15) we get:
\begin{equation}
\dot q_{_{1,2}}=\frac{\sqrt{\left| a+b\right|}}{2}
\sqrt{((q_{_{1,2}}^{2}-\phi _{_{1,2}}^{2})(q_{_{1,2}}^{2}-\Phi
_{_{1,2}}^{2}))}
\end{equation}

where

$$\theta
_{_{1,2}}=\dot
q_{_{1,2}}(0)+\frac{(a+b)}{4}(q_{_{1,2}}(0))^{4}+\frac{\omega
_{0}^{2}\pm \alpha}{2}\,(q_{_{1,2}}(0))^{2},$$

 $$\ \phi _{_{1,2}}=\frac{\omega _{0}^{2}\pm\alpha+
 \sqrt{(\omega_{0}^{2}\pm\alpha)^{2}-4\theta_{1,\,2}\left| a+b\right|}}{\left| a+b\right|},$$

$$ \Phi _{_{1,2}}=\frac{\omega _{0}^{2}\pm\alpha-
 \sqrt{(\omega_{0}^{2}\pm\alpha)^{2}-4\theta_{1,\,2}\left| a+b\right|}}{\left| a+b\right|}\,.$$

\bigskip

In (16) we have made assumption that $ a+b<0 $ because integration
of (17) can be done only in this case. The integration of (16)
gives
\begin {equation}
\frac{1}{\Phi _{_{1,2}}}F(\arcsin\frac{q_{_{1,2}}(t)}{\phi
_{_{1,2}}}\,;\,\frac{\phi _{_{1,2}}^{2}}{\Phi
_{_{1,2}}^{2}})=\frac {\sqrt{\left| a+b\right| }t}{2}
\end {equation}

where F(...) is the elliptic integral of the first kind [15]. Having
produced a reversion of the expression (17) finally we get:
\begin {equation}
q_{_{1,2}}(t)=\phi _{_{1,2}}sn(\frac{\sqrt{\left| a+b\right|
}}{2}\Phi _{_{1,2}}t\,;\frac{\phi _{_{1,2}}^{2}}{\Phi
_{_{1,2}}^{2}})
\end {equation}

where sn (....) is the elliptic sine of the Jacobi [15].

The expressions (18) are general solutions of the equations
corresponding to the Hamiltonian (13). By means of inverse
transformations $ x=\frac{q_{_{1}}+q_{_{2}}}{2}\, $;
$y=\frac{q_{_{1}}-q_{_{2}}}{2}\, $ they allow to express solution
of the initial problem in the analytical form through higher
transcendental functions. In case of the initial conditions $
\theta _{_{1,2}}=\frac{1}{4}\frac{(\omega _{0}^{2}\pm \alpha
)^{2}}{\left| a+b\right|} $

the solutions become considerably simpler [15] and takes form of solitary
waves [16]
\begin{eqnarray}
x=\frac{1}{2}(\sqrt{\frac{\omega_{0}^{2}+\alpha}{\left|a+b\right|}}
\tanh\sqrt{\frac{\omega_{0}^{2}+\alpha}{2}}t+
\sqrt{\frac{\omega_{0}^{2}-\alpha}{\left|a+b\right|}}
\tanh\sqrt{\frac{\omega_{0}^{2}-\alpha}{2}}t) \nonumber \\
y=\frac{1}{2}(\sqrt{\frac{\omega_{0}^{2}+\alpha}{\left|a+b\right|}}
\tanh\sqrt{\frac{\omega_{0}^{2}+\alpha}{2}}t-
\sqrt{\frac{\omega_{0}^{2}-\alpha}{\left|a+b\right|}}
\tanh\sqrt{\frac{\omega_{0}^{2}-\alpha}{2}}t)
\end{eqnarray}
For obtaining of solutions (20), (21) we neglected a dissipation of energy
and influence of external field, i.e. we suppose that the system (1) is
adiabatic isolated and change of energy during the one period of
oscillations can be neglected. Taking into account that the period of the
solution (18) is defined by the expression

$$T=4K(\frac{\phi _{_{1,2}}^{2}}{\Phi _{_{1,2}}^{2}})$$

(where K (...) is the complete elliptic integral [15]), it is possible to
announce that expressions (18) and (19) are valid for the intervals of time
\begin {equation}
t<\frac{\alpha\sqrt{(\omega _{0}^{2}\pm \alpha )}}{\Gamma}
4K(\frac{\phi _{_{1,2}}^{2}}{\Phi _{_{1,2}}^{2}})
\end {equation}

In case of intervals of time $ t>\frac{\alpha\sqrt{(\omega
_{0}^{2}\pm \alpha )}}{\Gamma}T $, under the action on the system
of external constant field the obtaining of the analytical
solution is impossible and it is necessary to use numerical
methods.

In case of strong nonlinearity for the particular values of parameters in
the system can occur chaotic oscillations [11-13] (see Fig 3, Fig 4, Fig 5).

Fig. 3 $x(t)\,\ $as a function of time obtained by numerical calculations
for the values of the dimensionless parameters: $\omega _{0}^{2}=10$, $\,\
\alpha =1$, $\,\ \Gamma =0.1$,\thinspace\ \ $f=1.7\,,\,\ \,a=3,\,\ \
b=2.5\,,\,\ \ c=2.5$

Fig. 4 as a function of time obtained by numerical calculations for the
values of the dimensionless parameters: $\ \ \ \ \ \ \ \ \ \ \ \ \ \ \ \ \ \
\omega _{0}^{2}=10$, $\,\ \alpha =1.2$, $\,\ \Gamma =0.1$,\thinspace\ \ $%
f=1.5\,,\,\ \,a=3,\,\ \ b=2.5\,,\,\ \ c=2.5$

Fig. 5 Projection of trajectory of phase point motion on the space $x(t)$%
,\thinspace\ $dx(t)/dt$,$\,\ y(t)$ \thinspace for the values of parameters:
\ $\omega _{0}^{2}=10$, $\,\ \alpha =1$, $\,\ \Gamma =0.1$,\thinspace\ \ $%
f=1.7\,,\,\ \,a=3,\,\ \ b=2.5\,,\,\ \ c=2.5\,\ \ $obtained by numerical
calculations.

It is ease to see from Fig. 3, Fig. 4 and Fig. 5 that motion is
realized in the restricted region of phase space. The external
field compensates energy dissipation and we have not focal points.
On the other hand motion is not periodically and as a result
trajectory is not closed (see Fig.5). In another words it means
that ratios between frequencies of oscillations of $ x(t),\,$
$\dot x(t),\,$ $ y(t)$ are not rational numbers. So in case of
strong nonlinearity, under the action of external constant field
in the system occurs chaotic non-periodical oscillations and
system is not integrable.
\newpage
References

1. Y.R.Shen '' The Principles of Nonlinear optics '', John Wiley and Sons,
New York, 1984

2. KitanoM, Yabuzaki T, Ogawa T, Phys. Rev Lett 1981 v.46 p926

3. Carmichael H.J., Savage C.M., WallsD.F. Phys. Rev. 1983

4. Kitano M., Yabuzaki T, Ogawa T., Phys. Rev. A., 1984 v 29

5. Yabuzaki T, Okamoto T., KitanoM., Ogawa T., Phys. Rev A. v29 p. 1964

6. Flitzanis C.L. Tang C.L. Phys. Rev. Lett. 1980 v.45. p 441

7. Goldstone J.A., Garmire E. Phys. Rev. Lett .1984 v.53, p.910

8. Rozanov N.N., Smirnov V.A., '' Pisma v JETP '', 1981 v.33 p. 504 (in
Russian)

9. Akhmanov S.A., Jeludeev N.I. izvestia of Ac.si of USSR, v.46 N.6, p.1070
1982 (in Russian)

10. Zheludeev N.I., Uspekhi Fizicheskix Nauk, v.157, N.4. p 685, 1989 (in
Russian)

11. Ott. E, '' Chaos In Dynamical Systems '' Cambridge, Univ. Press.
Cambridge 1993

12. Gutzwiller M.C., '' Chaos In Classical And Quantum Mechanics '',
Springer New-York 1990

13. Alligood K.T. Saner T.D. and York J.A. '' Chaos an Introduction to
Dynamical Systems '', Springer, New-York 1996

14. Lax P. Comm. Pure Appl. Math. v. 21 p.467 1968

15. Handbook of Math Functions, Edited by Abramovitz, National Bureau of
standards, Appl. 14. Math.series 55, Issued June 1964

16. R.Rajaraman, An Introduction to Solitons and Instantons in Quantum Field
Theory, North- Holland Publishing Company Amsterdam-New-York-Oxford, 1982

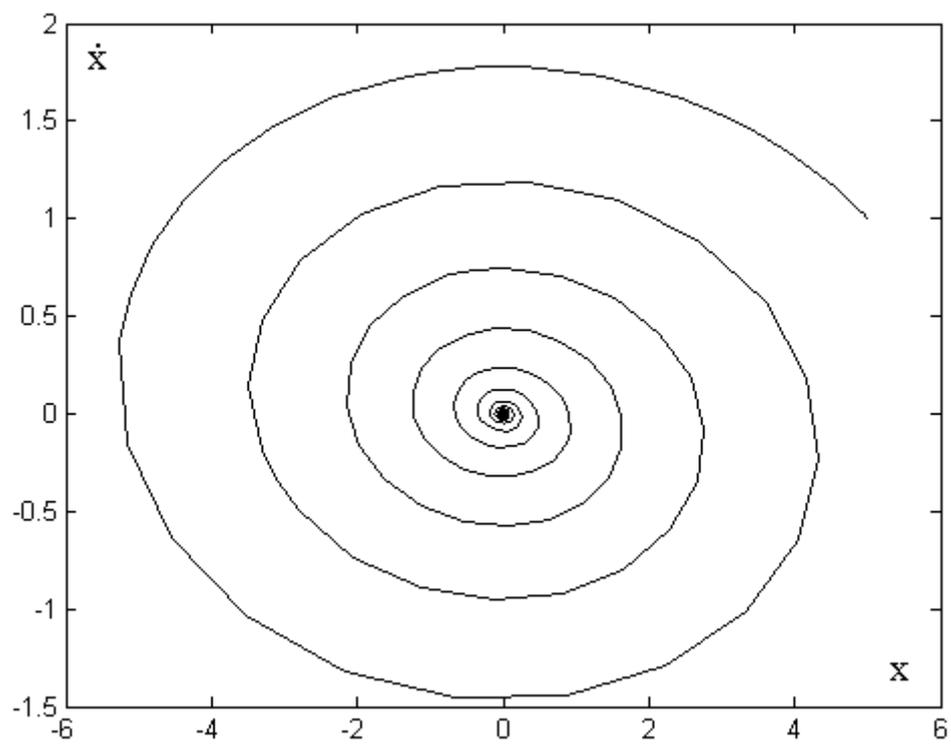
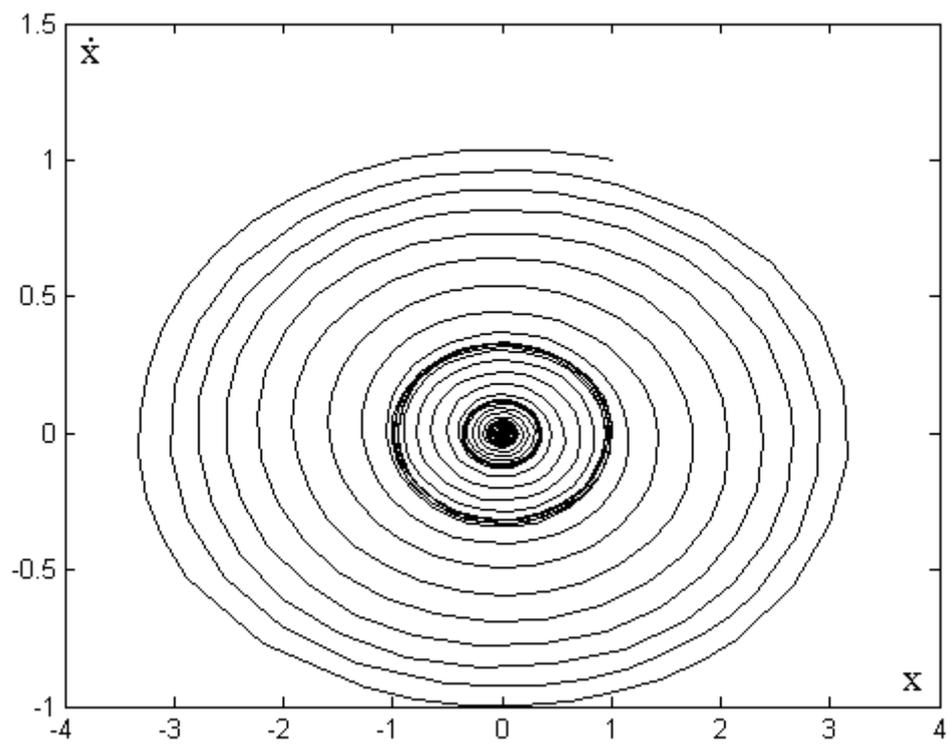
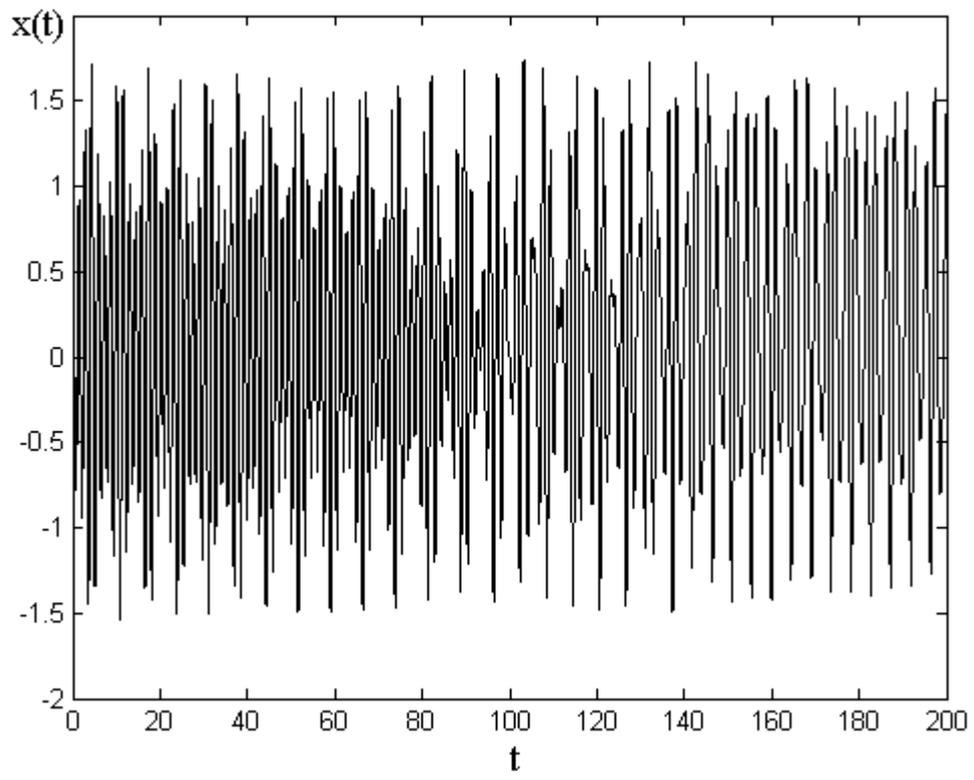
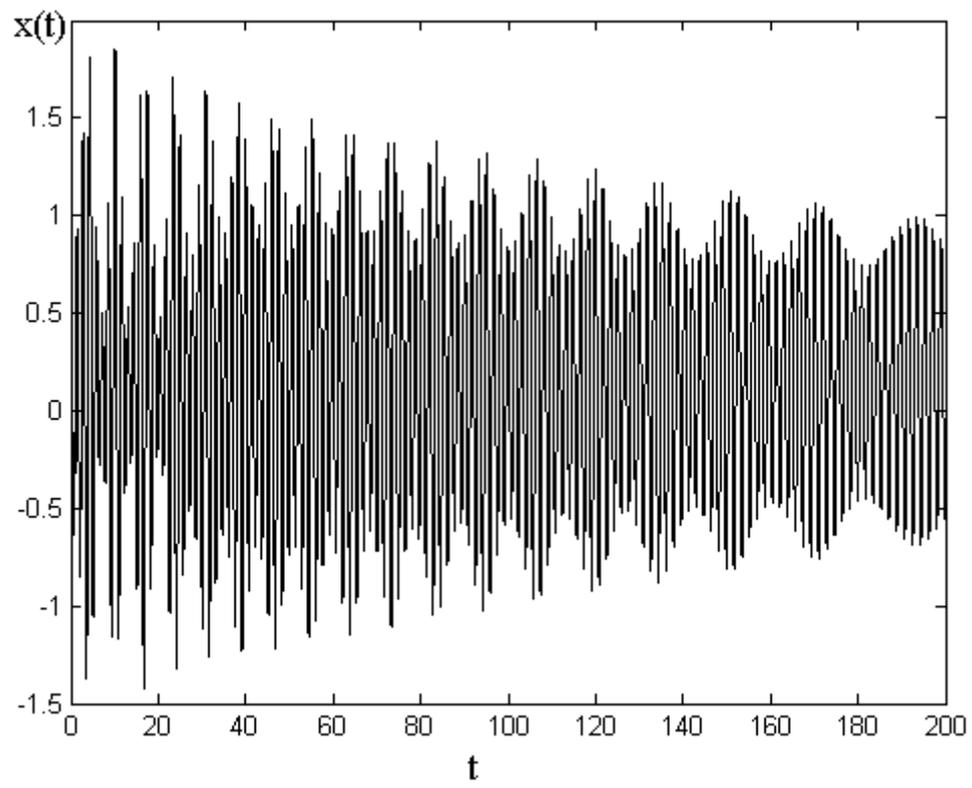
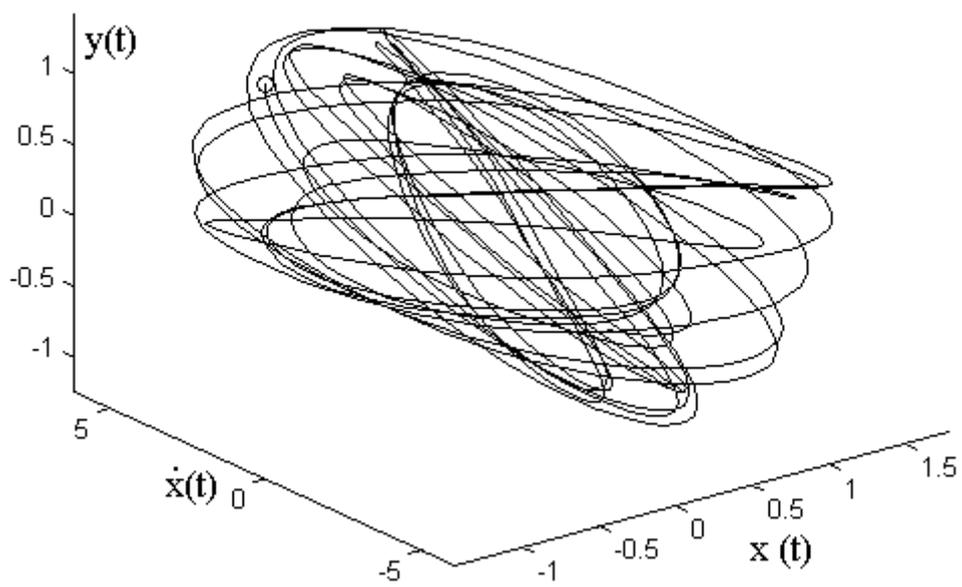
\end{document}